\documentclass[onecollarge]{svjour2}       
\usepackage{graphicx}
\usepackage{amssymb}
\usepackage{color}

\usepackage[UKenglish]{babel}

\begin{document}

\title{Incommensurability effects on dipolar bosons in optical lattices}

\author{Fabio Cinti}


\institute{Fabio Cinti \at 
             National Institute for Theoretical Physics (NITheP), Stellenbosch 7600, South Africa \\
             Institute of Theoretical Physics, Stellenbosch University, Stellenbosch 7600, South Africa \\
             \email{cinti@sun.ac.za}                 
}

\date{Received: date / Accepted: date}

\maketitle

\begin{abstract}

We present a study that investigated a quantum dipolar gas in continuous space where a potential lattice was imposed. 
Employing exact quantum Monte Carlo techniques, we analysed the ground state properties of the scrutinised system, varying 
the lattice depth and the dipolar interaction. For system densities corresponding to a commensurate filling with respect to the optical lattice, 
we observed a simple crystal-to-superfluid quantum phase transition, being consistent with the physics of dipolar bosons in continuous space.
In contrast, an incommensurate density showed the presence of a supersolid phase. 
Indeed, such a result opens up the tempting opportunity to observe a defect-induced supersolidity 
with dipolar gases in combination with a tunable optical lattice.
Finally, the stability of the condensate was analysed at finite temperature.



\end{abstract}

\section{Introduction}
\label{sec1}

During the last few years, tremendous development in the ability to control 
ultra-cold gases, characterised by long-ranged dipolar interactions, confined in optical lattices 
\cite{1367-2630-12-6-065025,PhysRevA.73.033605} has taken place.
In particular, surprising results have been achieved in quantum gases composed of 
Rydberg atoms \cite{Schaus:2015aa}, polar molecules \cite{Yan:2013aa} or lanthanides such as erbium (Er) \cite{1507.03500v1}. 

From a theoretical perspective, long-range magnetic or electric dipolar interactions are considered 
as chief candidates for observing and controlling novel phases in quantum many-body systems \cite{Baranov2008,Lahaye2009}.  
One of these phases is known as supersolidity, featuring both diagonal and off-diagonal order \cite{RevModPhys.84.759}.  
As is well known, the interaction among defects is a key ingredient for driving such an intriguing phase.  
Although in a quantum regime the defect interaction is still a problem not entirely understood in detail, 
it appears well established as, in some cases, supersolidity yields as a result of defects delocalisation \cite{Cinti:2014aa}.
Considering a classical regime only, some authors have emphasised that the effective defect interactions in a self-assembled crystal 
can be tuned from attractive to repulsive using an external periodic superlattice \cite{PhysRevE.88.060402}.

Concerning a quantum regime, recent studies have pointed out that the supersolid phase can be observed 
in optical lattices \cite{RevModPhys.84.759,PhysRevA.82.013645,PhysRevLett.95.237204,Pollet2010}.
For example, Pollet \textit{et al.} \cite{Pollet2010} have studied a complete phase diagram 
of a two-dimensional system composed of cold polar molecules on a triangular lattice.
The authors  proposed a phase diagram that featured a crystal, a superfluid, and, more important, a supersolid phase. 
However, even if  these studies have improved knowledge of the concept of supersolidity remarkably,
they still remain focused on single-band lattice models. 
Quantum phases using approximations that take into account  
higher bands still remain a subject to be understood in full \cite{PhysRevA.86.023617,PhysRevLett.104.090402}.

In this paper, we present the results concerning a boson dipolar system in continuous space where a potential lattice is imposed,
in other words removing the usual tight-binding condition. In such a continuous limit, the band structure is not completely formed, 
roughly depending on the potential depth. 

Using exact quantum Monte Carlo methods, we studied 
the many-body system investigating different lattice depths and dipolar interaction strengths, considering a 
filling factor around $n=1/3$. 
We showed that the behaviour of the dipolar systems in shallow lattices changed drastically considering a   
commensurate and an incommensurate filling of the lattice potential. 
In the first case one can observe a simple superfluid-to-crystal quantum phase transition, as already discussed in 
Refs.~\cite{PhysRevLett.113.240407,Buchler2007}.
Concerning the second situation, incommensurability features a defect-induced supersolidity, as originally proposed in Ref.\cite{Andreev1969}.  

The article is organised as follows: in the Section~\ref{sec2} we present the 
model Hamiltonian and the quantum Monte Carlo methods applied.
The results are presented and discussed in Section~\ref{sec3}. In particular,
Section~\ref{sec3a} refers to the ground state configuration, while Section~\ref{sec3b} 
is devoted to a regime of  finite temperature. 
Finally, in Section~\ref{sec4} a number of conclusions are presented.

\section{Model Hamiltonian and Methodology}
\label{sec2}

We considered an ensemble of bosons interacting via a dipole-dipole potential 
confined in a two-dimensional optical lattice. 
The system is described by a quantum-mechanical many-body Hamiltonian as follows
\begin{equation}
\label{hamiltonian}
{\cal H} =  \frac{\hbar^2}{2m}\sum_i \nabla^2_i + \sum_{i<j} \frac{D}{r_{ij}^3} - \sum_i u_i(\vec{r}),
\end{equation}
where $m$ is the mass of a single particle, while $r_{ij}=|\vec{r}_i - \vec{r}_j|$ is the distance between particles $i$ and $j$.
$D$ represents the characteristic strength of the interaction between two dipoles. 
The last term of the Hamiltonian refers to an optical triangular lattice potential
\begin{equation}
\label{oplat}
u(\vec{r}) =  u_0\left[\sin^2 \left( \frac{x+\sqrt{3} y}{2r_0} \right) + \sin^2 \left( \frac{- x + \sqrt{3}y}{2r_0} \right) +  
\sin^2 \left( \frac{x}{r_0} \right)\right]
\end{equation}
with $u_0$ being the lattice depth and $r_0$ the optical lattice constant.  
For the sake of clarity, we defined distances and energies in units of $d=D/r_0^3$. 

As mentioned before, 
the model proposed in equation (\ref{hamiltonian}) was investigated using a quantum Monte Carlo technique \cite{RevModPhys.67.279}.
More precisely, we sampled the density matrix of the system employing a path integral representation
in  continuous space. Our code was based on the well-known worm algorithm \cite{PhysRevLett.96.070601}. 
Over the last decade, this technique has been successfully implemented for studying the 
quantum properties of different bosonic systems \cite{0034-4885-75-9-094501}. 
An all-inclusive discussion of this methodology is provided in Ref.~\cite{Boninsegni2006}. 
Moreover, as pointed out lately \cite{Jain2011}, the employment of repulsive dipolar interactions combined with   
trapping potentials does not involve any particular pathology throughout the sampling stage.

The worm algorithm efficiently furnishes a numerically \textit{exact} estimation of the statistical observables, 
such as, for instance, energy per particle, density distributions and superfluid fraction. 
In accordance with Ref.~\cite{PhysRevB.36.8343}, we defined the last estimator as  
\begin{equation}
\label{sff}
f_S=\frac{m}{\hbar^2\beta N} \langle \vec{w}^{\,2} \rangle,
\end{equation}
where $\beta=1/k_BT$,  $\vec{w}=(w_x,w_y)$ being the winding number along the optical lattice in equation (\ref{oplat})in
and $N$ is the number of particles of the ensemble considered. We used up to $N=250$ particles and about 650 sites in order 
to exclude any finite-size effects. 
Even if the algorithm works at finite temperature, an accurate extrapolation of the ground state limit  (i.e. $T\to0$) 
may be reached as well. As a general rule, this limit is approached  when the 
structural and energetic properties of the system remain constant within the statistical error of the simulation.

The phase diagram of the Hamiltonian~(\ref{hamiltonian}) with $u_0=0$ has been intensively investigated over the last few years. 
At present, there is a general consensus that this phase diagram only presents a
melting quantum phase transition  from a triangular crystal to a homogeneous superfluid \cite{Buchler2007,PhysRevLett.113.240407,Astrakharchik2007}. 
The crystal-superfluid quantum phase transition results seem to be easily controlled by 
adjusting the dimensionless parameters $r_d=Dm/a\hbar^2$, $a$ being the averaged interparticle distance 
of the purely continuous system. B\"uchler \textit{et al.} have identified the transition at $r_d^{QM}=18(4)$ \cite{Buchler2007}. 
Nevertheless, some aspects related to the transition order continue to be controversial. 
In particular, as discussed by Spivak and Kivelson \cite{PhysRevB.70.155114}, 
at the interface between the two phases, the system should feature a microemulsion phase.
Using a variational quantum Monte Carlo \cite{Sarsa}, 
Moroni and Boninsegni have recently pointed out that ``for all practical purposes" \cite{PhysRevLett.113.240407},
the transition results of the first order, with a coexisting phase (microemulsion) mainly inaccessible for any possible experiment. 

Differently from the case just discussed, the limit $u_0\to\infty$ presents a more complex but also more interesting phase diagram. 
Here, between a superfluid and a crystal phase, Pollet \textit{et al.} \cite{Pollet2010} identified a supersolid phase 
as well as a microemulsion as proposed in Ref. \cite{PhysRevB.70.155114}. The authors found that for a commensurate filling factor $n=1/3$   
($n$ typically being the ratio between simulated particles and lattice size), the system displayed 
diagonal and off-diagonal long-range order concurrently.  

In order to observe a supersolid phase, in this study  we were  interested in analysing the unexplored limit of  finite  $u_0$,  
considering again the  filling values $n=1/3$ (commensurate filling) and $n\gtrsim$1/3 (noncommensurate filling).
The non-commensurate filling case was studied by introducing an interstitials density ranging from 0.02 to 0.04. 
In addition, we focued our attention on the limit $r_d\lesssim r_d^{QM}$, in other words far from a free-space triangular lattice phase.

\section{Results}
\label{sec3}

\subsection{Ground state properties}
\label{sec3a}

\begin{figure}[t]
\centerline{\includegraphics[height=.5\textwidth]{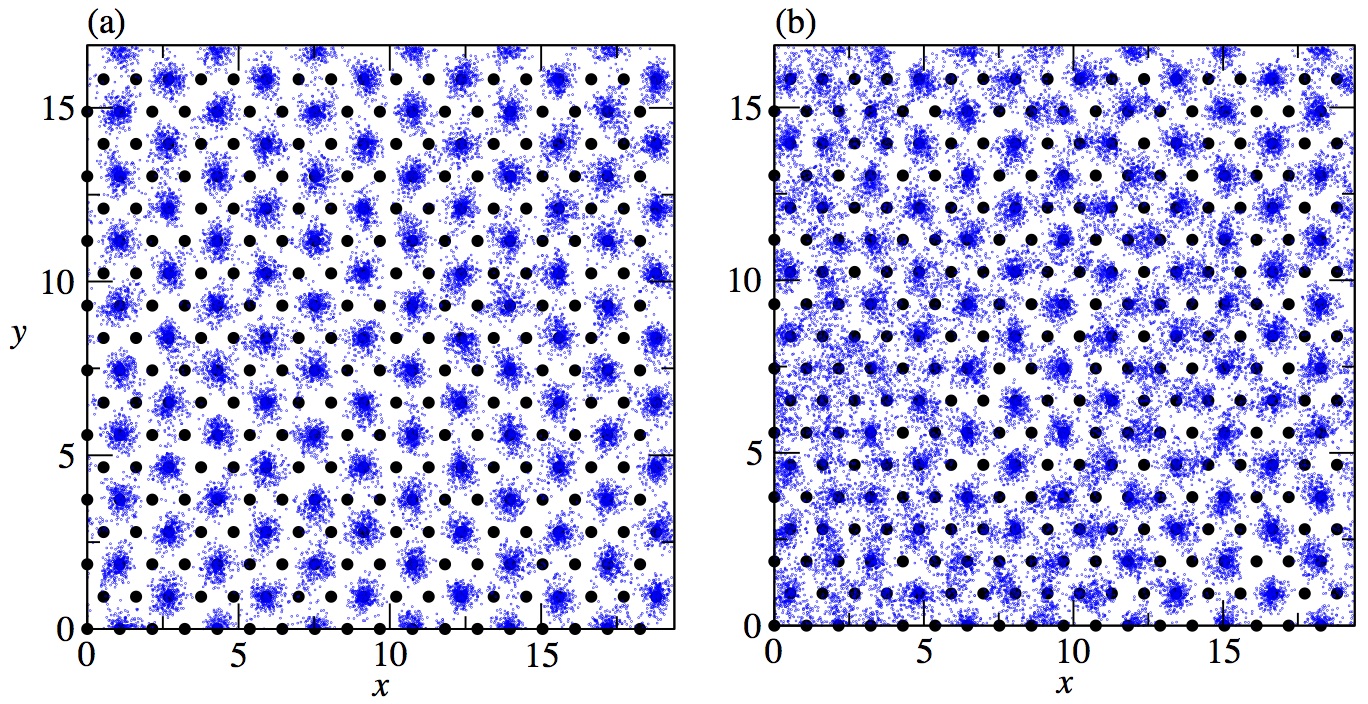}}
\caption{Example of configuration snapshots, in other words  particle world lines projection on an $xy$-plane, for a system of dipolar bosons 
confined in a two-dimensional optical lattice with equation~(\ref{oplat}). The dots represent the 325 sites, in other words the potential minimum. 
The interaction is $d=15$ and the lattice depth is $u_0=8$.  (a) $N=108$ (no defects), and (b) $N=120$ (density defect 0.036).}
\label{cofings}
\end{figure}

Figure~\ref{cofings} depicts snapshots of the projection of world lines onto the $xy$-plane for $n=1/3$ (Figure~\ref{cofings}a),
with a density of defects (interstitials) equal to 0.036 (Figure~\ref{cofings}b). The dipole-dipole interaction is $d=15$ (i.e. $r_d\approx$5)
while the lattice depth corresponds to $u_0=8$.
The representation in Figure~\ref{cofings} provides a functional way to sketch out
the probability distribution of the many-body system in real space \cite{RevModPhys.67.279}. Below a superfluid transition temperature, 
the overlap of paths entails exchanges among bosons and superfluidity too. 

We observe in Figure~\ref{cofings}a that particle paths are completely confined around the local 
minima of the lattice potential (\ref{oplat}), showing a stripe crystal ground state configuration. 
The situation changes if one introduces defects. Figure~\ref{cofings}b depicts 
a configuration whereby a  localised path occurs with delocalised ones. This coexistence implies the presence of 
a supersolid phase.

\begin{figure}[t!]
\centerline{\includegraphics[height=.45\textwidth]{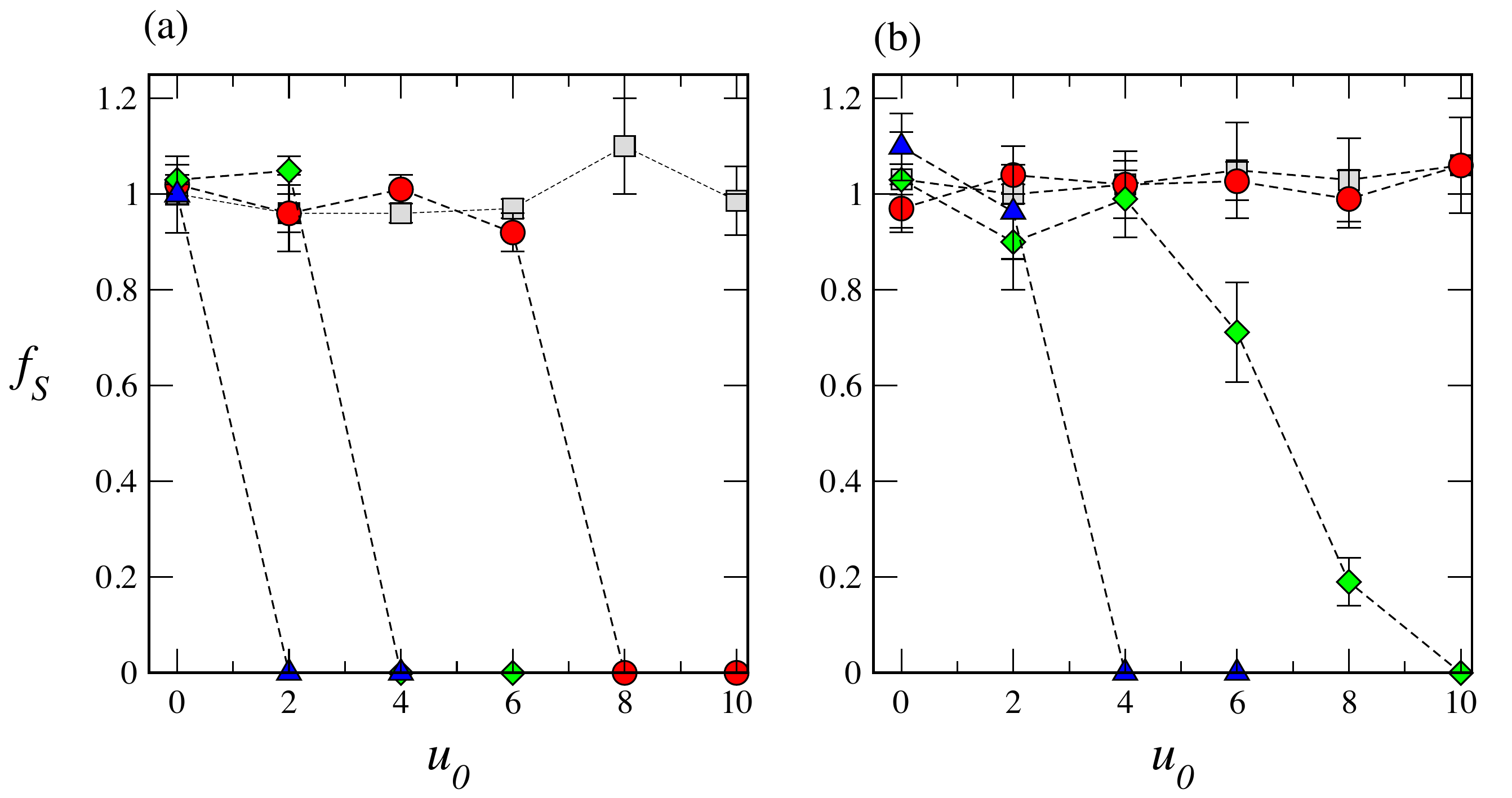}}
\caption{Superfluid fraction versus lattice depth considering a system without defects (a) and introducing a defect density equal to 0.036.
$r_d = 1.67$ (square), 3.34 (circle), 5 (diamond) and 6.67 (triangle).}
\label{superfluid}
\end{figure}

Figure~\ref{superfluid} shows the superfluid fraction $f_S$ versus $u_0$, considering the different values of $r_d$,
again for a commensurate (Figure~\ref{superfluid}a) and an incommensurate (Figure~\ref{superfluid}b) sample, respectively.
Analysing Figure~\ref{superfluid}a, we see that only for $r_d=1.67$ the dipolar system persists in being superfluid ($f_S=1$) over 
all the $u_0$ considered. However, by increasing  $r_d$ we obtain a simple drop of the superfluid estimator from one 
(homogeneous superfluid)
to zero (crystal) when the lattice depth turns deeper. Such behaviour signalises a simple superfluid-to-insulating-crystal quantum phase transition. 
The snapshot configuration in Figure~\ref{cofings}a is a representative
example that shows how the lattice is forcing a crystal phase onto the system.  
In fact,  for $u_0=0$ we have observed a superfluid phase for all the $r_d$ considered. 

Figure~\ref{superfluid}b shows simulations introducing defects, again for different $r_d$.
Here for $r_d=3.34$ and $u_0\leq10$,  we notice that the presence of defects seems to  
remove only the transition, leaving the superfluid phase unchanged.
Moreover, for interaction strengths $r_d = 1.67$ and 6.67, $f_S$ does not show any fundamental 
changing with respect to the commensurate case.  
Yet, for $r_d=5$ and $5<u_0<9$, the ground state superfluidity 
leads to a $nonhomogeneous$ superfluid behaviour, characterised by  $0<f_S<1$ \cite{PhysRevLett.25.1543}.
This set of parameters allows the system to move into a supersolid phase, as shown in Figures~\ref{cofings}b.

\begin{figure}[t]
\centerline{\includegraphics[height=.4\textwidth]{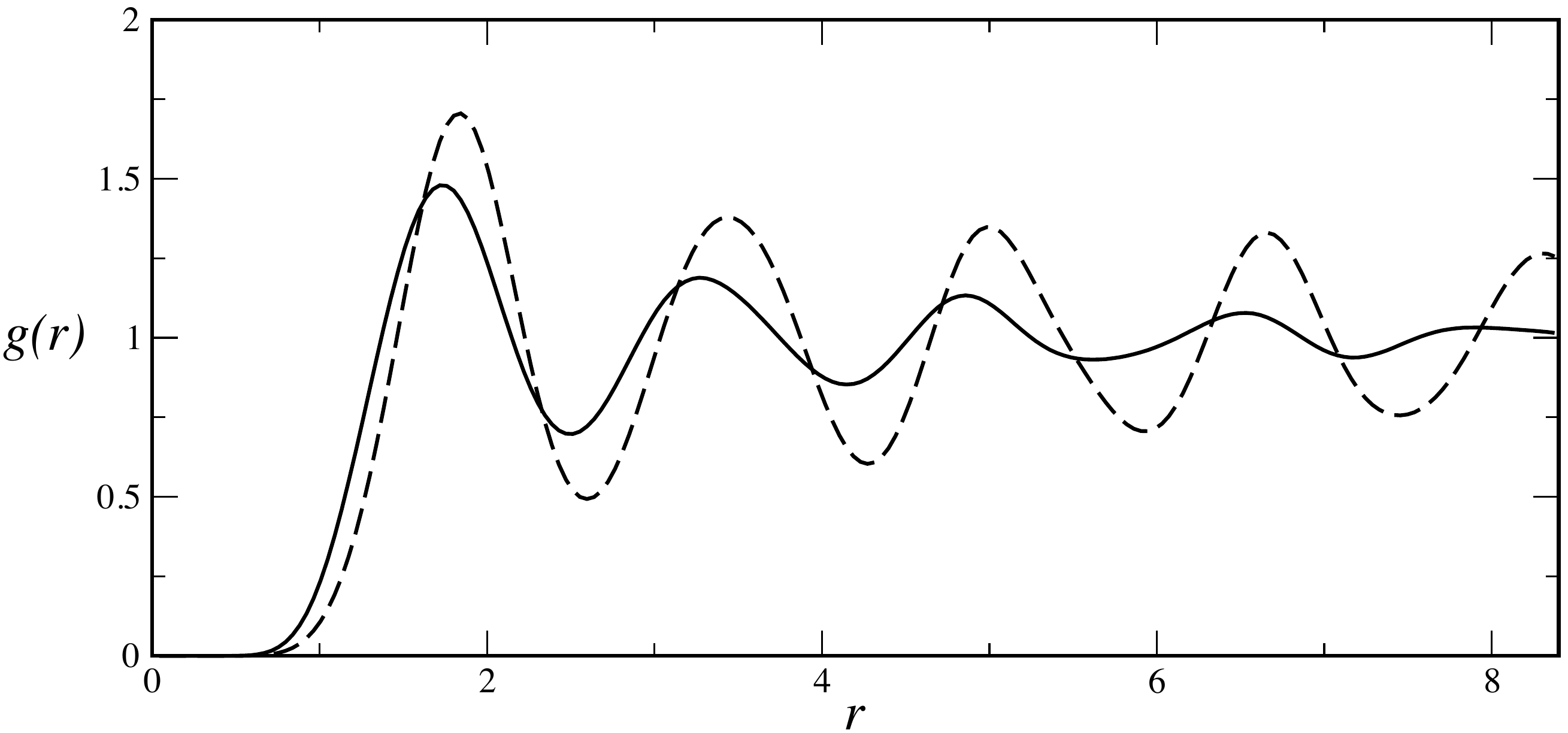}}
\caption{Pair correlation function $g(r)$ for a crystal phase (dashed line) and a supersolid phase (continuous line), using the same set of parameters as in Figure~\ref{cofings}.}
\label{gr}
\end{figure}

Figure~\ref{gr} reports the two-point density correlation function $g(r)$. 
Such a function provides more qualitative information on a liquid or crystal phase.
The figure still compares a commensurate (dashed line) and an incommensurate
(continuous line) filling value, respectively. 
In either case the system perfectly mimics the lattice periodicity, with the first maximum 
representing the averaged interparticle distance for a stripe crystal.
Regarding the supersolid phase (continuous line), $g(r)$ still presents a robust modulation 
that becomes smoother at large distances due to a strong particle delocalisation throughout the lattice.

\subsection{Finite temperature properties}
\label{sec3b}

\begin{figure}[t]
\centerline{\includegraphics[height=.33\textheight]{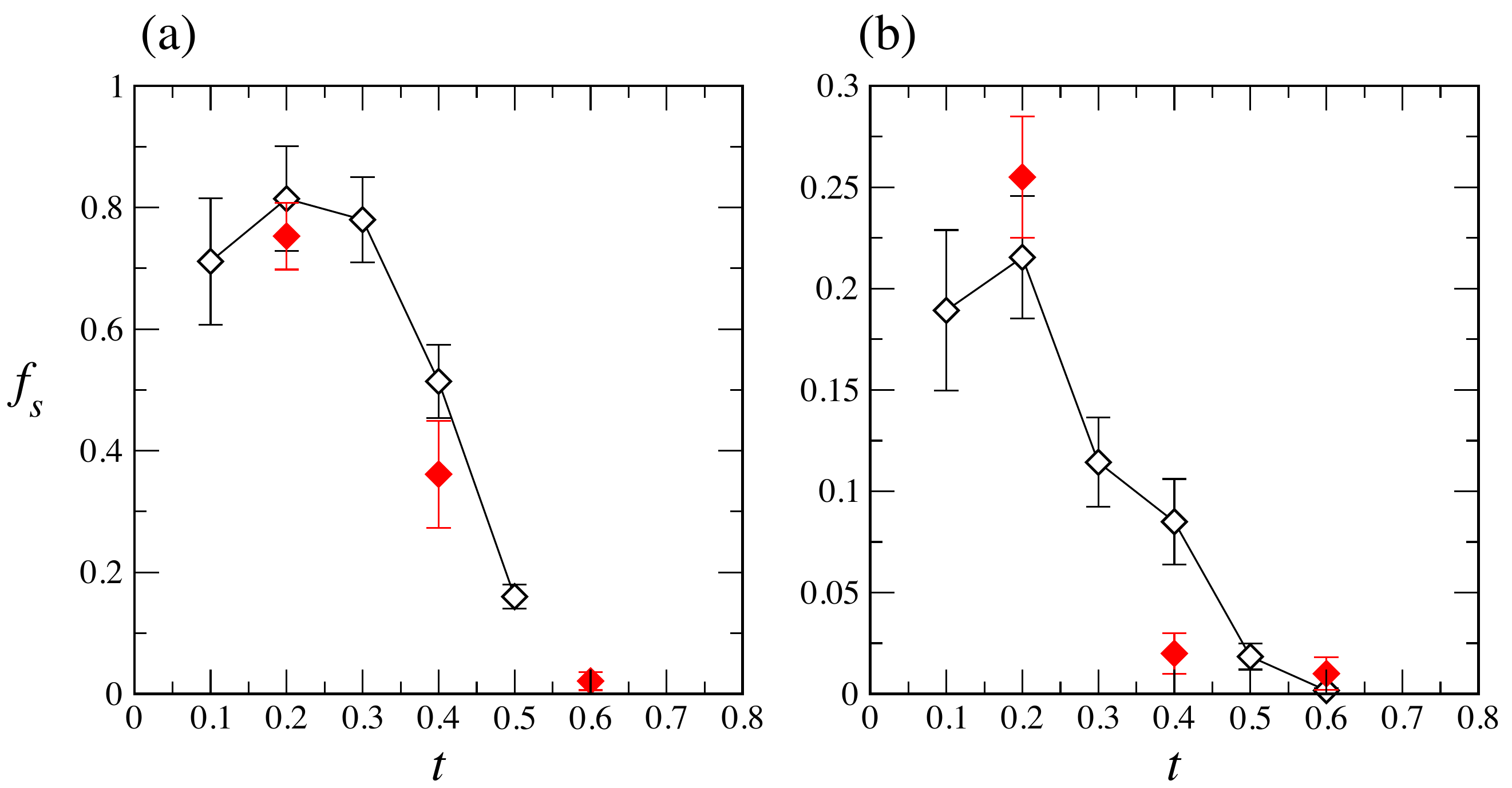}}
\caption{Superfluid fraction versus the reduced temperature $t$ with a defect density equal to 0.036 considering $u_0=6$ (a) and $u_0=8$ (b).
$N=120$ (open diamond) and 212 (full diamond).}
\label{fst}
\end{figure}

Now we discuss the temperature effects on the supersolid phase observed in Figure~\ref{cofings}b and~\ref{superfluid}b. 
As one would expect, the stability of the condensate at finite temperature is a key feature for supporting realistic experiments.
In order to clarify the behaviour in temperature, 
Figure~\ref{fst}  depicts the superfluid fraction as a function of temperature,
for $u_0=6$ (Figure~\ref{fst}a) and 8 (Figure~\ref{fst}b), respectively. 
In accordance with Figure~\ref{superfluid}b, we take into account a dipole strength ($r_d=5$) 
that furnishes a nonhomogeneous superfluidity for $t\to0$.
We defined a reduced temperature $t\equiv k_{B}T/  (\hbar^2 n/m) $, 
$\hbar^2 n/m$ being the kinetic energy at the mean interparticle distance.  

Considering first $u_0=6$ for an ensemble of $N=120$ dipoles,
one sees that $f_S$ remains constant ($f_S\sim$0.75) up to $t\lesssim0.4$,
revealing a critical behaviour at finite $t$ (temperature phase transition). 
This, therefore, appears as entirely consistent with the
Berezinsky-Kosterlitz-Thoules (BKT) theory for a two-dimensional system with a continuous symmetry \cite{chaikin2000principles}. 
It is well known that for a homogeneous gas of bosons, the transition 
temperature can be  estimated  at $T_{BKT}=2\pi\hbar^2n/mk_B$.
Figure~\ref{fst}b shows that this approximation yields a BKT regime for  $t\lesssim 0.85$. 
The lowering of the BKT regime in temperature is strictly connected with the interaction strength 
showed in Figure~\ref{fst}, consistent with the results discussed in Ref.~\cite{PhysRevLett.105.070401}.
Figure~\ref{superfluid}b shows a similar physics for $t\lesssim0.3$ 
but with a lower superfluid fraction in the ground state ($f_S\sim$0.22).
We observe that in the supersolid region, $u_0$ appears to influence $t_{BKT}$ only mildly.
This, actually, is an interesting feature of dipolar bosons in an optical lattice. 
For $u_0>8$, we do not observe any superfluidity (and consequently any $t_{BKT}$), 
leading to the onset of a crystal phase. These results seem to be consistent with a first-order phase transition.
It is worthwhile stressing that a crystal-to-supersolid  first-order phase transition has been observed also for 
ultra-cold soft-core bosons \cite{PhysRevA.87.061602,1367-2630-16-3-033038} and 
for dipolar bosons in triangular lattices \cite{PhysRevA.85.021601} as well. 

Finally, in order to exclude finite size effects, Figure~\ref{fst} compares two different 
system sizes, $N=120$ (open diamond) and $N=212$ (full diamond). It clearly appears that superfluidity at finite 
temperature does not change within the statistical errors for both sizes. 

\section{Conclusions}
\label{sec4}

In this study we considered a two-dimensional dipolar bosonic gas in the presence of a weak triangular optical lattice. 
The results were obtained using an exact quantum Monte Carlo that implements the worm algorithm in continuous space. 
Different from previous studies, we investigated a Hamiltonian~(\ref{hamiltonian}) modifying the depth of the optical lattice (\ref{oplat}) and the 
strength of the dipole-dipole interactions. 
Regarding a commensurate filling, we observed that the presence of a periodic potential 
did  not change the phase diagram of the system for $u_0=0$, in other words where the quantum gas simply shows a crystal-to-superfluid quantum phase 
transition.  
In contrast, the introduction of defects into the system was found to allow a clear defect-induced supersolidity. 
We  also verified that this phase remained thoroughly  solid even at finite temperature.  
Finally, the supersolid mechanism here discussed fully agrees with the original definition of supersolid phase given by Andreev and Lifshitz \cite{Andreev1969}.
In this limit of density, our results  therefore proved that the experimental realisation of this long-sought quantum phase
can also be made using quantum dipolar gases in optical lattices.

\begin{acknowledgements}
The author thanks  G. Pupillo and T. Macr\`i for enlightening discussions.
\end{acknowledgements}





\end{document}